\documentstyle[12pt]{article}

\begin{document}

\begin{center}

{\Large{\bf Crystal Symmetry and Time Scales} \\ [5mm]
V.I. Yukalov$^1$ and E.P. Yukalova$^2$} \\ [3mm]
{\it $^1$Bogolubov Laboratory of Theoretical Physics \\
Joint Institute for Nuclear Research, Dubna 141980, Russia \\
and \\
$^2$ Laboratory of Informational Technologies \\
Joint Institute for Nuclear Research, Dubna 141980, Russia}

\end{center}

\vskip 2cm

\begin{abstract}

The relation between the notion of crystalline symmetry and
characteristic time intervals when this symmetry could be observed
is analyzed. Several time scales are shown to exist for a system
of interacting particles. It is only when the observation time is
much larger than the mesoscopic fluctuation time, the notion of
crystalline symmetry becomes physically meaningful. The ideas are
concretized by a two-phase lattice model.

\end{abstract}

\vskip 2cm

\section{Characteristic Time Scales}

Crystalline symmetry is a well-defined geometric notion. But a
crystal as such is a physical object consisting of many moving
particles. In what sense these particles form a crystalline
lattice with a given symmetry? An intuitive answer to this question
would be that it is the average positions of particles which form
a crystalline lattice, with the averaging performed over
sufficiently large time. But, to be mathematically correct, it is
necessary to concretize what does mean "sufficiently large" with
respect to the observation time. The aim of this report is to
demonstrate that the minimal time scale providing the lower
threshold for the observation time is the mesoscopic fluctuation
time.

There are several characteristic time scales for a system of
interacting particles [1]. The smallest typical time is the {\it
interaction time}
\begin{equation}
\label{1}
\tau_{int} \sim\; \frac{a_s}{v} \; ,
\end{equation}
in which $a_s$ is an effective interaction radius, or effective
particle size, or scattering length, and $v$ is a characteristic
particle velocity, say, sound velocity. When observing a system of
particles during an {\it observation time} $\tau_{obs}$ that is
smaller than $\tau_{int}$, the motion of these particles looks
absolutely random, with no preferable positions in space.

Another important scale is the {\it local-equilibrium time}
\begin{equation}
\label{2}
\tau_{loc} \sim\; \frac{\lambda}{v}\; ,
\end{equation}
where $\lambda$ is a mean free path. The latter is connected with
the scattering length $a_s$ and the average density of particles,
$\rho$, as $\lambda\sim(\rho a_s^2)^{-1}$. The average density
$\rho\sim a_0^{-3}$, with $a_0$ being a mean interparticle
distance. Hence, $\lambda\sim a_0^3/a_s^2$, from where
\begin{equation}
\label{3}
\tau_{loc} \sim\left ( \frac{a_0}{a_s}\right )^3 \tau_{int} \; .
\end{equation}
During the local-equilibrium time, particles interact with each
other, so that the state of local equilibrium is being evolved.

At a very short time $t$, one has the {\it dynamic stage}, when
\begin{equation}
\label{4}
0< t< \tau_{int} \; ,
\end{equation}
and particles move randomly. No spatial symmetry in the positions
of particles can be noticed for the observation time in the
diapason (4). The second time interval corresponds to the {\it
kinetic stage}, when
\begin{equation}
\label{5}
\tau_{int} < t <\tau_{loc} \; .
\end{equation}
At this stage, interparticle correlations begin arising due to
interactions, but the overall motion looks yet rather
disorganized, resembling the Brownian motion. In solids, $a_0$ is
just slightly larger than $a_s$, so that the local-equilibrium
time (3) is only a little longer than the interaction time,
$\tau_{loc}\sim\tau_{int}$. The kinetic stage (5) is practically
absent in solids. One has the {\it hydrodynamic stage}, when
\begin{equation}
\label{6}
\tau_{loc} < t < \infty \; .
\end{equation}
It is at this stage when the averaged motion of particles is
commonly assumed to define a crystalline lattice symmetry, under
the appropriate external conditions.

However, in general, it is not the local-equilibrium time (3) that
gives the lower threshold for the observation time, in order that
the lattice symmetry be well defined. The actual lower threshold
is much larger than $\tau_{loc}$. There exists one more time scale
that can be called the {\it mesoscopic fluctuation time}
\begin{equation}
\label{7}
\tau_f \sim\; \frac{l_f}{a_0} \; \tau_{loc} \; ,
\end{equation}
in which $l_f$ is a linear size of a quasiequilibrium fluctuation
[1]. The fluctuation is termed mesoscopic because of its
intermediate size, being in the interval $a_0\ll l_f\ll L$, where
$L$ is a characteristic length of the system. Since $l_f\gg a_0$,
one has $\tau_f\gg \tau_{loc}$.

In this way, the hydrodynamic stage (6) consists of two
qualitatively different parts. One is the {\it mesoscopic stage},
when
\begin{equation}
\label{8}
\tau_{loc} < t <\tau_f \; .
\end{equation}
Here, crystalline symmetry cannot yet be rigorously defined for
the whole system, although the main part of the latter may
already be approximately described by a kind of symmetry. At the
{\it macroscopic stage}, when
\begin{equation}
\label{9}
\tau_f < t <\infty \; ,
\end{equation}
the averaged motion of particles can be correctly characterized by
assigning a concrete spatial symmetry. It is the mesoscopic
fluctuation time (7) that defines the lower threshold for the
observation time which is necessary for
making the notion of symmetry physically meaningful. In this
regard, sufficiently large observation time implies
\begin{equation}
\label{10}
\tau_{obs} \gg \tau_f \; .
\end{equation}
Only averaging over the observation time satisfying the inequality
(10) provides a mathematically correct definition of the related
lattice symmetry.

\section{Two-Phase Lattice Model}

In order to examplify the existence of mesoscopic fluctuations,
let us consider a lattice model for which such fluctuations not
only do exist but their presence makes the system more stable.
Lattice models describe the physics of solids where an essential
concentration of vacancies can arise. These vacancy defects appear
in crystals with high anharmonicity. Among such objects, we could
mention quantum crystals [2--6], crystallized plasma [7],
crystallized white dwarfs [8], crystals near liquid-solid
transition [1,9,10] and crystals close to vitrification [1,11].
Mesoscopic fluctuations of heterophase nature are common for
liquid crystals [12].

It is worth noting that mesoscopic heterophase fluctuations exist
not only in crystals but also in other materials. Thus, they are
clearly observed by neutron scattering experiments in many
magnetic alloys [13,14]. Even in such classical ferromagnets as Fe
and Ni, there appears local magnetic order in their paramagnetic
phase, which is also a sort of mesoscopic fluctuations [15,16].
Similar phenomenon occurs in high-temperature superconductors
where it is often termed phase separation [17--22]. More examples
can be found in Ref. [1].

Now, let us turn to the lattice model to be analyzed in what
follows. Consider a system of $N$ particles in volume $V$. Each
particle can occupy one and only one lattice site of a given
crystalline lattice. Not all lattice sites are occupied, but some
of them are vacant. The distribution of vacancies through the
volume is not uniform. Some parts of the system contain more
vacancies than other, which results in essentially different
density of particles in the corresponding regions. The location
and shapes of these parts with different density are not fixed,
but fluctuate in space and time, forming a system with mesoscopic
density fluctuations [23]. Thus, we have a heterophase system with
coexisting dense and dilute phases. Such a system can serve as a
cartoon of a crystal with regions of disorder [24].

Let us label the quantities related to the dense phase with the
index 1 and those corresponding to the dilute phase, by the index
2. For instance, the average number of particles pertaining to the
dense phase is $N_1$, occupying the volume $V_1$, while the average
number of particles in the dilute phase is $N_2$, inside the volume
$V_2$. The total number of particles and the system volume are
conserved:
\begin{equation}
\label{11}
N=N_1 + N_2 \; , \qquad V = V_1 + V_2 \; .
\end{equation}
Because of the presence of vacancies, the number of lattice sites,
$N_0$, is larger than the total number of particles $N$. Hence,
the density of sites is larger than that of particles:
\begin{equation}
\label{12}
\rho_0 >\rho \; , \qquad \rho_o \equiv \frac{N_0}{V} \; , \qquad
\rho\equiv \frac{N}{V} \; .
\end{equation}
By definition, the dense phase has a larger density of particles
than the dilute phase:
\begin{equation}
\label{13}
\rho_1 >\rho_2 \; , \qquad \rho_\nu \equiv \frac{N_\nu}{V_\nu}
\qquad (\nu=1,2) \; .
\end{equation}
Introducing the geometric probabilities of the corresponding
phases,
\begin{equation}
\label{14}
w_\nu \equiv \frac{V_\nu}{V}\; , \qquad w_1 + w_2 = 1\; ,
\end{equation}
the mean density of particles can be written as
\begin{equation}
\label{15}
\rho= w_1\rho_1 + w_2\rho_2 \; .
\end{equation}
For what follows, it is convenient to define the dimensionless
densities
\begin{equation}
\label{16}
n_\nu \equiv \frac{\rho_\nu}{\rho_0} \; , \qquad
n\equiv \frac{\rho}{\rho_0} =\frac{N}{N_0} \; ,
\end{equation}
for which Eq. (15) reduces to
\begin{equation}
\label{17}
n=w_1 n_1 + w_2 n_2 \; .
\end{equation}

A system with mesoscopic heterophase fluctuations is
quasiequilibrium and has to be described by a quasiequilibrium
Gibbs ensemble [1], with a statistical operator $\rho(\xi)\sim
e^{-Q(\xi)}$, where $Q(\xi)$ is a quasihamiltonian and the set
$\xi=\{ \xi_\nu({\bf r})\}$ of manifold indicators characterizes a
distribution of phases in space. This phase distribution is frozen
at the mesoscopic stage (8), when no lattice symmetry can be
defined. Going to the macroscopic stage (9) implies averaging over
all phase configurations. This averaging can be defined [1] as
functional integration over the set $\xi=\{\xi_\nu({\bf r})\}$ of
the manifold indicator functions. In the process of that
integration, the notion of the {\it heterophase Hamiltonian}
$\overline H$ arises, defined by the relation
\begin{equation}
\label{18}
\int \; e^{-Q(\xi)}\; {\cal D}\xi
\equiv e^{-\beta\overline H} \; ,
\end{equation}
in which $\beta\equiv T^{-1}$ is inverse temperature, $k_B\equiv
1$. From Eq. (18), one has
\begin{equation}
\label{19}
\overline H = - T\ln \int e^{-Q(\xi)}\; {\cal D}\xi \; .
\end{equation}

The $i$-site of the lattice, pertaining to the part filled by the
$\nu$-phase, is characterized by the variable $e_{i\nu}$, taking
the value $e_{i\nu}=1$ if the site is occupied by a particle and
the value $e_{i\nu}=0$ when the site is empty. For the heterophase
Hamiltonian (19), we obtain
\begin{equation}
\label{20}
\overline H = H_1\oplus H_2 \; ,
\end{equation}
which is a sum of the {\it phase-replica Hamiltonians}
\begin{equation}
\label{21}
H_\nu = w_\nu \; \sum_{i=1}^{N_0} \left ( K_i -\mu\right ) e_{i\nu}
+
\frac{1}{2}\; w_\nu^2 \; \sum_{i\neq j}^{N_0} \Phi_{ij}\; e_{i\nu}
e_{j\nu} \; ,
\end{equation}
where $K_i$ is kinetic energy; $\mu$, chemical potential; and
$\Phi_{ij}$ is interaction potential. The phase probabilities
$w_\nu$ are defined from the minimization of the thermodynamic
potential
\begin{equation}
\label{22}
y=-\; \frac{1}{N}\; \ln\;{\rm Tr}\; e^{-\beta\overline H} \; ,
\end{equation}
in which, for taking account of the normalization $w_1+w_2=1$, it
is convenient to introduce
\begin{equation}
\label{23}
w\equiv w_1 \; \qquad w_2\equiv 1 - w\; .
\end{equation}
Then $w$ is given by the equations
\begin{equation}
\label{24}
\frac{\partial y}{\partial w} = 0 \; , \qquad
\frac{\partial^2 y}{\partial w^2} > 0 \; .
\end{equation}
In terms of the occupation operators $e_{i\nu}$, the densities
(13) are
\begin{equation}
\label{25}
\rho_\nu =\frac{1}{V} \; \sum_{i=1}^{N_0} <e_{i\nu}>\; ,
\end{equation}
where $<\ldots>$ implies the statistical averaging, and the
densities (16) become
\begin{equation}
\label{26}
n_\nu =\frac{1}{N_0} \; \sum_{i=1}^{N_0} < e_{i\nu}> \; .
\end{equation}
Recall that, by definition, $\nu=1$ corresponds to the dense phase
while $\nu=2$, to the dilute phase, so that $\rho_1>\rho_2$ and
$n_1>n_2$.

It is possible to pass to quasispin representation by introducing
the variable $\sigma_{i\nu}=\pm 1$, defined by the relations
\begin{equation}
\label{27}
\sigma_{i\nu} \equiv 2e_{i\nu} -1 \; , \qquad e_{i\nu} =
\frac{1}{2}\left ( 1 +\sigma_{i\nu}\right ) \; .
\end{equation}
Then the Hamiltonian (21) takes the form
\begin{equation}
\label{28}
H_\nu = U_\nu + \frac{1}{8} \; w_\nu^2 \; \sum_{i\neq j}^{N_0}
\Phi_{ij}\sigma_{i\nu}\; \sigma_{j\nu} - \;
\frac{1}{2}\; w_\nu \; \sum_{i=1}^{N_0} h_{i\nu}\sigma_{i\nu}\; ,
\end{equation}
where
$$
U_\nu \equiv \frac{1}{2}\; w_\nu \; \sum_{i=1}^{N_0} \left (
K_i -\mu\right ) + \frac{1}{8} \; w_\nu^2 \; \sum_{i\neq j}^{N_0}
\Phi_{ij} \; ,
$$
$$
h_{i\nu} \equiv \mu - K_i - \; \frac{1}{2}\; w_\nu\;
\frac{1}{N_0} \; \sum_{i\neq j}^{N_0} \Phi_{ij} \; .
$$
The form (28) corresponds to the Ising model in an external field.

To accomplish concrete calculations, let us consider the Kac-type
interaction potential $\Phi_{ij}=\Phi_{ij}(N_0)$, for which
\begin{equation}
\label{29}
\lim_{N_0\rightarrow\infty} \Phi_{ij}(N_0) = 0 \; , \qquad
\lim_{N_0\rightarrow\infty} \left | \frac{1}{N_0} \;
\sum_{i\neq j}^{N_0} \Phi_{ij}(N_0)\right | < \infty \; .
\end{equation}
For this kind of potentials, the problem as is known [25], can be
solved asymptotically exactly in the thermodynamic limit, when
$N_0\rightarrow\infty, \; V\rightarrow\infty$, so that
$N_0/V\rightarrow const$. In the case of such long-range
potentials, the mean-field decoupling
\begin{equation}
\label{30}
<\sigma_{i\nu}\;\sigma_{j\nu}>\; =\delta_{ij} +
(1-\delta_{ij})<\sigma_{i\nu}><\sigma_{j\nu}>
\end{equation}
becomes asymptotically exact.

\section{Numerical Results and Conclusions}

With the decoupling (30), we can calculate all thermodynamic
characteristics of the model. In terms of the occupation operators
$e_{i\nu}$, this decoupling is equivalent to the equality
$$
<e_{i\nu}\; e_{j\nu}>\; = \frac{1}{2}\; \delta_{ij}
(<e_{i\nu}>\; + \; < e_{j\nu}>) + (1 -\delta_{ij})\; < e_{i\nu}>
<e_{j\nu}> \; .
$$
Calculations result in a system of transcendental equations which
we have analyzed numerically. Dimensionless densities
$n_1 \equiv a, \; n_2 \equiv b, \; (a>b)$,
as functions of temperature $T$, in units of
$\Phi\equiv \frac{1}{N_0} \; \sum_{i\neq j}^{N_0}\Phi_{ij}$,
are presented in Fig. 1, for different mean densities. Note that
from the condition of heterophase stability, that is the second of
Eqs. (24), it follows that $\Phi>0$. Fig. 2 shows the probability
of the dense phase vs. temperature. The shifted chemical potential
$\mu_*$, defined by the equations
$\mu_* \equiv (\mu-K)/\Phi,\; K\equiv
\frac{1}{N_0}\; \sum_{i=1}^{N_0}\; K_i$,
is given in Fig. 3. The pressure, $P$, specific heat, $C_V$, and
isothermal compressibility, $\kappa_T$, are shown in Figs. 4 to 6,
respectively. All curves in Figs. 1 to 6 end at the points where
the probability of the dense phase becomes zero, after which the
heterophase system looses its stability, since the phase
probabilities, by definition, must be always positive. Finally, in
Fig. 7, the thermodynamic potential $\omega_w \equiv yT/\Phi$
is drawn for the case of the heterophase system and also this
potential, $\omega_1$, for the pure dense system when $w=1$. As is
seen, one always has $\omega_w<\omega_1$, which means that the
heterophase system is thermodynamically more stable than the pure
system.

In this way, a heterophase system, exhibiting mesoscopic
fluctuations, is thermodynamically more stable than a pure system
without these fluctuations. At the same time, the existence of
mesoscopic fluctuations implies the occurrence of local
instability [1]. In the present case, such instabilities are
related to the fluctuational grouping of vacancies in some parts
of the system, which forms fluctuating disordered regions locally
destroying crystalline symmetry. It is only for large observation
times, satisfying the inequality (10), when the system displays,
on average, a perfect symmetry of the lattice, while for times
shorter than the lifetimes of mesoscopic fluctuations, an ideal
lattice symmetry cannot be observed. However, such a {\it temporal
breaking of symmetry} makes the system more stable in the long
run.

The results examplified here by a lattice model can be generalized
to other physical systems. To our mind, the same conclusions
concern not only simple physical systems but, even to a greater
extent, all complex systems, such as societies. Thus, a society
allowing the occurrence of mesoscopic fluctuations can be called
democratic, while that one prohibiting any deviation from the
prescribed symmetry is dictatorial. In the long run, {\it
democracy is more stable than dictatorship}.

\newpage

\begin{center}
{\bf Figure Captions}
\end{center}

{\bf Fig. 1}. The density of the dense phase (solid curve) and of
the dilute phase (dashed curve) as functions of temperature.

\vskip 5mm

{\bf Fig. 2}. The probability of the dense phase vs. temperature.

\vskip 5mm

{\bf Fig. 3}. The shifted chemical potential vs. temperature.

\vskip 5mm

{\bf Fig. 4}. The dimensionless pressure vs. temperature.

\vskip 5mm

{\bf Fig. 5}. Specific heat vs. temperature.

\vskip 5mm

{\bf Fig. 6}. Isothermal compressibility vs. temperature.

\vskip 5mm

{\bf Fig. 7}. Dimensionless thermodynamic potentials for the
heterophase system (two lower curves) and for the pure dense
system (two upper curves).

\vskip 5mm

\end{document}